\documentclass[sigconf,authorversion,nonacm]{acmart}

\usepackage{multirow}
\usepackage{microtype}

\AtBeginDocument{%
  \providecommand\BibTeX{{%
    \normalfont B\kern-0.5em{\scshape i\kern-0.25em b}\kern-0.8em\TeX}}}

\setcopyright{acmcopyright}
\copyrightyear{}
\acmYear{}
\acmDOI{}

\acmConference[BIR 2024]{14th International Workshop on Bibliometric-enhanced Information Retrieval}
\acmPrice{00}

\begin{document}

\title{Enhancing Research Information Systems with Identification of Domain Experts
}


\author{Gautam Kishore Shahi \& Oliver Hummel}
\affiliation{%
  \institution{University of Applied Sciences, Mannheim}
  \country{Germany}
    }
    \email{o.hummel@hs-mannheim.de}





\renewcommand{\shortauthors}{Anonymous Author(s)}

\begin{abstract}
Research organisations and their research outputs have been growing considerably in the past decades. This large body of knowledge attracts various stakeholders, e.g., for knowledge sharing, technology transfer, or potential collaborations. However, due to the large amount of complex knowledge created, traditional methods of manually curating catalogues are often out of time, imprecise, and cumbersome. Finding domain experts and knowledge within any larger organisation, scientific and also industrial, has thus become a serious challenge. Hence, exploring an institution's domain knowledge and finding its experts can only be solved by an automated solution. This work presents the scheme of an automated approach for identifying (scholarly) experts based on their publications and, prospectively, their teaching materials. Based on a search engine, this approach is currently being implemented for two universities, for which some examples are presented. The proposed system will be helpful for finding peer researchers as well as starting points for knowledge exploitation and technology transfer. As the system is designed in a scalable manner, it can easily include additional institutions and hence provide a broader coverage of research facilities in the future.

%

\end{abstract}


\maketitle

\section{Introduction}

In recent years, on the one hand, research institution and their research output have become more visible due to the advancement of scholarly databases, data-sharing policies, and willingness to collaborate amongst research institutions \cite{farooq2023knowledge}. Nevertheless, most research institutions still follow the traditional approach of merely listing metadata (such as titles and author names) of research results on their websites alongside hand-curated profiles containing usually rather coarse-grained areas of expertise. Hence, these websites usually only provide vague and often outdated information about researchers and especially their specific expertise. This poses a major challenge for stakeholders interested in understanding the research landscape or looking for domain experts or knowledge in, e.g. a nearby institution.

The number of research institutions has roughly doubled in every decade, and the number of researchers has increased in a similar fashion \cite{dong2017century}. Most institutions, however, are lagging in updating publication metadata for their researchers, which leads to reduced visibility for researchers as well as existing knowledge and hence limits the value of research institutions as nuclei for innovation, especially for the surrounding regional industrial ecosystem. This situation is especially unpleasant for small and medium-sized companies that cannot afford dedicated scientific staff who are able to screen and penetrate scientific literature. Moreover, current research information management systems (RIMS) are not capable of automatically keeping track of research areas that are tackled by publications. Consider a researcher who started in the field of Natural Language Processing and Information Retrieval and later also started working on chatbots and large language models (LLM) as an example. Due to the involved manual work, RIMSs or websites are often not updated in a timely manner with such evolving research domains. If, in this example, someone would be looking for an expert in artificial intelligence language technology, then a RIMS might not give accurate results due to its rather static content. Moreover, even the use of RIMS might still not be widely adopted, as it is often the case for smaller Universities of Applied Sciences in Germany, where research has only slowly been gaining importance in recent years. Hence, interested stakeholders usually have to rely on open-domain search engines to find helpful experts from nearby research institutions, which in turn have to rely on the rather static web content of these institutions.

With the advance of Generative Artificial Intelligence \cite{strobel2024exploring}, various AI systems, such as ChatGPT or Gemini, have become publicly available and, according to previous research, might be a promising solution for this challenge \cite{askari2023test}. However,  after asking ChatGPT the question "\textit{Can you please provide a list of domain experts in the field of big data at Hochschule Mannheim?}", it merely replied the following: \\
"\textit{As of my last update in January 2022, I don't have access to specific lists of domain experts at Hochschule Mannheim (...); I recommend visiting the university's website, department pages, or contacting the relevant faculty or research centres directly. They can provide you with information about faculty members, researchers, and experts in the field of big data at Hochschule Mannheim.}"

This example anecdotally illustrates that even the most advanced open-domain chatbots are currently over-challenged with this specific exercise and merely refer the user to perform a standard web search. To overcome such time-consuming manual searches of domain experts based on static and potentially not updated information, we propose a knowledge-based search engine, which is able to automatically extract the research field(s) of scientists based on their publications and other information published on the Web and, prospectively, also on materials from internal learning management systems, such as Moodle. The key contributions of this paper are as follows, it presents:
\begin{itemize}
    \item ideas for extracting the field of research from scientific articles.
    \item a search engine for finding domain experts based on the field of research
    \item a prototypical version of the proposed system. 
\end{itemize}
In the remainder of this paper, we briefly discuss the state of the art in section~\ref{sec:related} and the proposed approach itself in section~\ref{sec:method}. After that, we discuss its implementation in section~\ref{sec:implementation} and some preliminary results in section~\ref{sec:result}. Finally, we present important ideas for future work in section~\ref{sec:future} and conclude our work with section \ref{sec:conclusion}.

\section{State of the Art}
\label{sec:related}

As the literature illustrates, the topic of classifying scientific articles into their respective field of research is still emerging. Until today, academic institutions have mostly used a manual approach for collecting and analysing scholarly data \cite{guest2013collecting}. For example, while reviewing the research data management at his institution, the author of \cite{perrier2017research} was confronted with the fact that data is still collected manually to deliver simple services such as a list of publications per researcher. Hence, it is not possible to search for researchers based on a given topic. The research data in the university of the author of \cite{schuetzenmeister2010university} was also curated manually; overall this is time-consuming and produces delays in collecting and publishing the data. The same holds true for the University of the authors, where publication lists are still managed with the help of Excel tables and not centrally published at all.

Another interesting study that has been conducted on the information seeking behaviour of users of 17 search systems for academics has found that these search systems basically use very simple keyword searches and hence bear great potential for improvement through more advanced search functionalities \cite{nedumov2019exploratory}. In the study reported by \cite{schopf2023knowledge}, an exploratory search using semantic technologies is used to provide better access to domain experts. However, it is merely based on the research area provided by the researchers and manually fed into the system. In another study, the author proposes REDI, a Linked Data-powered framework for managing and storing academic data \cite{ortiz2022redi}. However, they also still use static data that is provided manually by researchers for this purpose.

Beyond purely theoretical research, now there are also some workshops and challenges emerging, aiming at building and comparing models able to classify research contained in scientific publications, such NSLP 2024\footnote{https://nfdi4ds.github.io/nslp2024} for example: This exchange and such challenges are likely to attract different approaches and models for classification in the future and, hence, help with the advancement of the scientific community in this important field.

\section{Approach}
\label{sec:method}
Given the insights from, e.g., \citep{nedumov2019exploratory}, it is clear that a purely search-based solution for research information systems will be as imperfect as other manually curated catalogues, such as in libraries. Since the development of a complex research system is a highly dynamic process that -- according to experience from various fields such as software engineering, design thinking, or entrepreneurship -- needs to be user-centric, we adopt a highly agile approach with rapid prototyping and brief feedback cycles mainly inspired by the Lean Startup method \cite{ries2011lean}.

For the current early stage of our prototype, we envisage a research interested person searching for domain experts as our central persona and attributed two use cases to it, namely directly finding domain experts based on a classic keyword search and identifying the domains of expertise that are actually represented at an institution. The main idea is to implement a prototypical solution according to the following scheme and, once this is accomplished, evaluate it with researchers and other colleagues involved in research management and technology transfer at our university.

\begin{figure}
  \centering
  \includegraphics[width=0.49\textwidth]{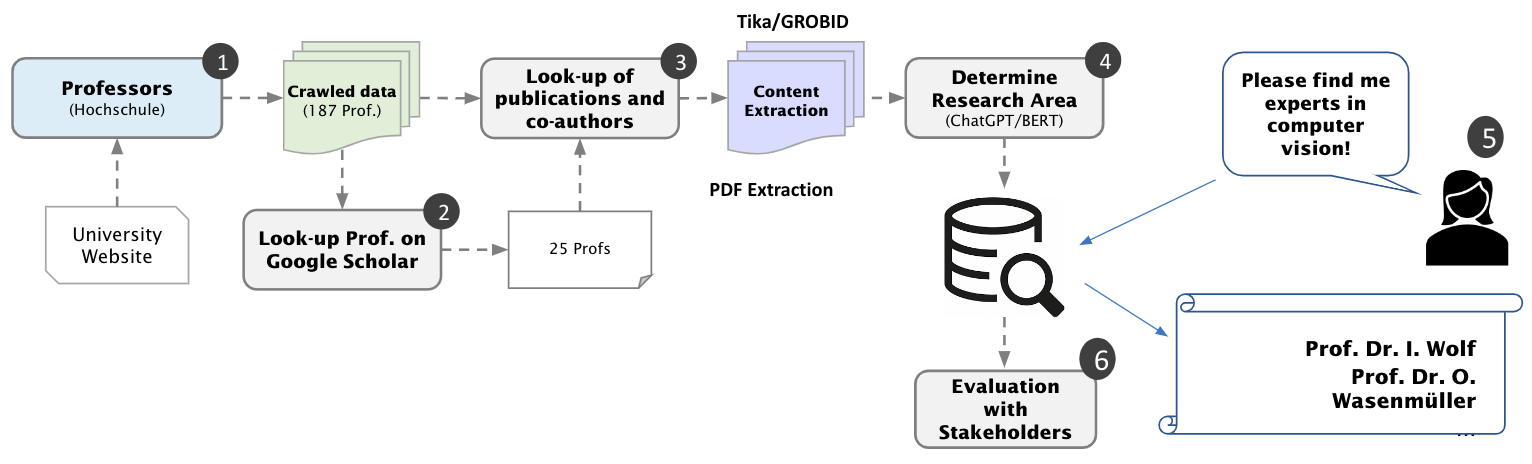}
\caption{Flow diagram for proposed improved Research Information System.}
  \label{fig:pipeline}
\end{figure}

In the approach designed so far (cf. Figure~\ref{fig:pipeline} for an illustration), we aim to ingest a given list of researchers from a university or a similar organisation in order to avoid noisy data usually coming from a general web crawl. With an official list provided by the university, one can crawl for publications, e.g., via the university's homepage or scrape it from another source, such as Research Gate or Google Scholar. Another advantage of using an officially provided list is that, in our case (and probably most other cases as well), it contains at least some helpful metadata, such as the department or the broader subject area. For the actual crawling, we apply a heuristic approach, for which we, e.g., take the university name or the email's domain into account to get the best possible matches. Once a researcher's name is identified with a given degree of certainty, we crawl their information, such as citations over past years, co-authors, and lists of publications, and try to find the PDFs of the publications on the Web where possible.

Once the papers are extracted, we extract their content from the PDFs \cite{lin2024revolutionizing}. For the time being, we use ChatGPT's API to identify a research area for the extracted content \cite{okey2023investigating}, which is then added to the corresponding researcher's profile. Both the texts of the crawled materials, i.e., the papers, as well as the researcher profile, are then stored in a search engine as described in the section immediately following, where we also describe the other steps involved in this process in more detail. In the section that follows the next one, we elaborate on how we plan to further enhance classic search technology to achieve better results in the quest for domain experts in the future.

\section{Implementation}
\label{sec:implementation}
This section discusses a prototypical implementation of our proposed approach with data from Hochschule Mannheim University of Applied Sciences\footnote{https://www.english.hs-mannheim.de/the-university.html}. Currently, the proposed system is implemented based on a list of professors by the university. Below, we highlight the most important insights from implementing the sketch from Figure~\ref{fig:pipeline}:

\textbf{Gathering Professors}. First, we parsed a list of professors from the university website. In total, it contained 188 professors that are listed together with some metadata like department, email, and telephone number. 

\textbf{Crawling Publication Data}. Based on this initial list of professor names, we used a crawling script in Python using an open-source\footnote{https://pypi.org/project/scholarly/} library to search for the name of each professor on Google Scholar. If there was a match, we extracted metadata of professors, such as given research areas, citation counts, and list of publications for a total of 28 professors, most of them coming from the departments of computer science and biotechnology. These 28 matches were manually verified for correctness. 

\textbf{Collecting Publications}. For each professor, we scraped a list of publications using Beautifulsoup \cite{richardson2007beautiful} to gather additional publication information like title, author, and link to the PDF of the paper. Currently, we have links for 420 publications and were able to download 268 of them for our analysis. The remaining papers were not avaiblable to us, mainly because they were behind a paywall or otherwise not accessible. Once the PDFs were downloaded, we used another Python script to parse the PDF using TIKA\footnote{https://pypi.org/project/tika/} and GROBIRD\footnote{https://github.com/kermitt2/grobid} to extract the textual content of the paper, excluding references as this might add unnecessary noise to searches later. After cleaning out further unwanted information, like email ID or URLs, we indexed the texts in our search engine. 

\textbf{Identifying Research Areas}. To identify the research area of each professor, we are currently evaluating three approaches. First, we used the metadata from the university homepage; second, we scraped the data entered by the researchers themselves in Google Scholar. Third, we aim to extend these by extracting more fine-granular information from each downloaded paper with the help of a Large Language Model (LLM), as indicated before. Currently, we do this by simply calling the ChatGPT API, providing it with the research area classification from the Library of Congress \cite{salaba2023cataloging}, and asking it to classify the field of research for each given paper accordingly. For each author, we merged the research areas delivered by ChatGPT for his papers with those provided by the university, as well as with those retrieved from Google Scholar. The result gives a relatively broad overview of the research expertise of each professor. 

\begin{figure}
    \centering
    \includegraphics[width=0.46\textwidth]{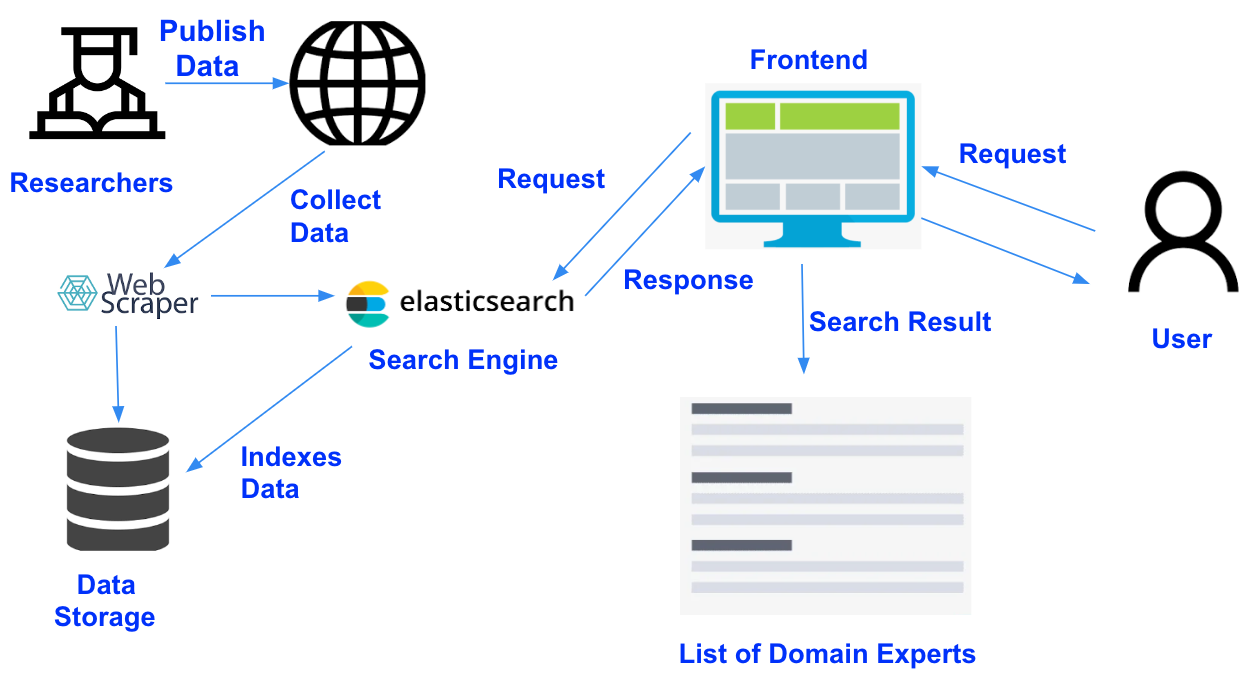}
\includegraphics[width=0.32\textwidth]{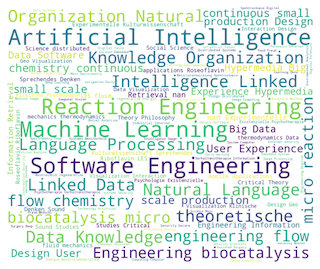}

    \caption{(a) Architecture diagram and data flow for search experience (b) Word cloud generated from the research areas of professors with Google Scholar profiles at Hochschule Mannheim }
    \label{fig:archictecure_wordcloud}
\end{figure}

\textbf{Illustrating Research Areas}. From the union of all extracted research areas, we derived a word cloud using bi-gram tokens. An example is shown in Figure~\ref{fig:archictecure_wordcloud} (b).

This word cloud is intended to get stakeholders interested in a university an up-to-date overview of research topics that are currently addressed at an institution.

\subsection{Search Engine} 

We are currently using Elasticsearch 8.7.0\footnote{https://www.elastic.co/guide/en/elasticsearch/reference/current/release-notes-8.7.0.html} as our core search engine since it provides out of the box text search functionality as well as advanced text analysis features for the data we collect. The overall architecture, data model, and search \& browse interface are discussed in the following.

\textbf{Architecture}. We propose a proof of concept for our system based on a classic client-server architecture. The backend consists of a Flask application \cite{shahi2022mitigating}, which operates as a server and is 

connected to an Elasticsearch index containing all collected data. The front end is a client-side single-page application also based on Flask with a "thin server" architecture. I.e., most business logic is moved from the server to the client that requests data when needed, thereby allowing for a seamless user experience. This architecture is shown in Figure~\ref{fig:archictecure_wordcloud} (a), which also explains the flow of data triggered by a user request. 

\textbf{Data Model}. The data model of the application is based on the information required from a researcher. Hence, it consists of the data extracted from the university website, Google Scholar, and the research areas extracted from scholarly publications. The data model can later be discretionarily extended to support further entities and their associated data. 

\begin{figure}
  \centering
  \includegraphics[width=0.49\textwidth]{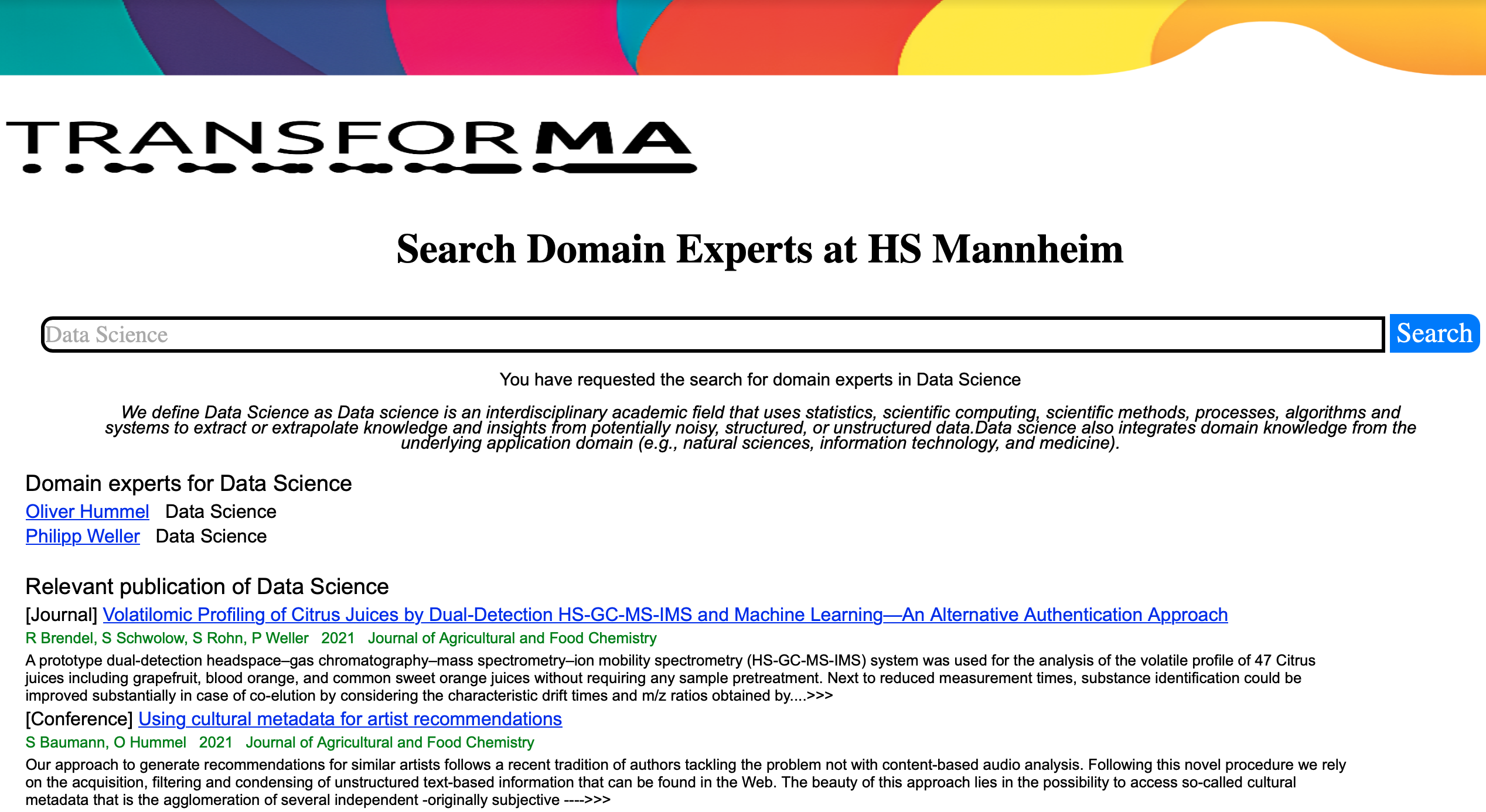}
  \caption{Exemplary Search Results with domain experts and relevant publications.}
  \label{fig:search}
\end{figure}

\textbf{Search \& Browse}. The homepage of the search engine initially shows the word cloud of the research areas available in the institution to provide an overview of the fields of expertise that are present there. An information-seeking stakeholder can then start looking for the desired experts by entering a search term (i.e., a desired research area) in the text box, or they can browse a sortable list of research fields that serve as a starting point to get to a domain expert. Once the search button is clicked, the user gets a basic definition of the search term extracted from Wikipedia, as well as a list of available domain experts. The user can further click on the domain expert to get more detailed profile information, such as on the research area, publications, and potential links to other bibliographic sources. A glimpse of the search interface with the interface is shown in Figure~\ref{fig:search}.

\section{Preliminary Results and Lessons Learned}
\label{sec:result}

We have implemented the approach as described before, and since search has become a fairly well-understood topic in recent years, we have experienced no unexpected issues with its basic functionality.
What we have learned so far is that manually attributed research areas are typically more coarse-grained than those that are extracted by ChatGPT, so it seems like a good complement at first glance. Even after a closer look, the research areas extracted from ChatGPT make sense and nicely illustrate how this approach is able to highlight even recent trends in a personal publication history. Consider the following research areas extracted for one professor at our university:
\begin{itemize}
    \item \textbf{University Website}: Big Data, Data Science, Information Retrieval, Software Engineering
    \item \textbf{Google Scholar}: Big Data, Software Engineering, Information Retrieval
    \item \textbf{Extracted from publications by ChatGPT}: Cognitive Neuroscience, Software Design Patterns, Object-Oriented Programming
\end{itemize}

The ChatGPT data is apparently fine-grained and also mirrors a very recent collaborative work of this colleague in an unusual field. However, it is, of course, reasonable to question whether one joint publication in an area like "Cognitive Neuroscience" turns a computer scientist into an expert in psychology. Thus, it might make sense to consider further metrics, such as the number of papers in an area or something similar, to obtain even better results in the future.

Another lesson learned is that although a word cloud seems to be a nice visual aid at first glance, our current implementation reveals some weaknesses at second glance. First and foremost, it is visible that not all research areas are well represented by bi-grams. Moreover, a mixture of languages (such as English and German in our case) in non-English speaking countries might be somewhat confusing for prospective users and needs to be fixed in future versions.

\section{Future Work}
\label{sec:future}

We plan to extend the search database to a partner university to add more researchers and demonstrate that our approach is generalisable to multiple institutions in the future. We have also planned to integrate additional data sources, like other academic search engines or platforms such as Research Gate or DBLP, to increase the coverage of our approach. Moreover, it is necessary to improve the quality of the word cloud since, obviously, not all research areas are well represented by bi-grams. One way to handle this better might be to use a positive list of research areas as a filter for retrieved results. As mentioned before, it also seems necessary to add some basic language detection and translation capabilities so that a word cloud does not contain a mixture of various languages, such as English and German, as in our example. In the short term, we are also aiming to improve the design of the search page by adding more functionality, such as a chatbot for answering questions about the domain and providing contact details. 

In the spirit of the Lean Startup approach, we also plan to gather feedback from potential users of our system to make sure that we are actually developing a useful piece of technology. We also plan to use semantic web technologies for the mapping of research areas within ontologies \cite{nandini2019ontology,dutta2015mod}.

Another possible more long-term extension for our work is to test different (and locally hosted) LLM implementations for the extraction of subject areas from papers and to evaluate the results obtained from them. As the preliminary results from ChatGPT illustrated, it still seems necessary to better understand the accuracy of the search results in general and the applicability of LLMs for such tasks in particular. 

The knowledge embodied in research publications is certainly important and valuable; however, it is probably only one side of the coin as it mostly covers the latest research results. As most scientists also have teaching obligations, it is probably safe to assume that a large part of their more fundamental knowledge is embodied in teaching materials. This is obviously less interesting for research transfer, but nevertheless, it might be interesting for finding potential teachers for advanced training, e.g., technical domains, and, of course, for broadening and sharpening the recognised knowledge areas of domain experts.

However, teaching materials are usually not published and, hence, not freely available. With access to the course management platform (CMP) of an institution, which should be possible for our system once it becomes officially used there, downloading these materials is probably less a technical issue and more a question of Copyright and willingness of the affected colleagues to at least share their materials for analysis with our system, if they do not want their scripts and slide decks to become publicly available. Hence, integrating a "stealth" mode for files that should be analysed but not indexed, as well as a crawler for our institution's CMP (which is Moodle) into our system, is another future task we are about to tackle soon.

\section{Conclusion}
\label{sec:conclusion}

Finding domain experts at research institutions is a challenging task that current search- or even catalogue-based approaches are not able to achieve. Hence, in this work, we presented the core of a modern research information system that is able to extract the research field from scientific publications and can be searched using a web interface. Built upon an open-source search engine, it can provide a list of domain experts for a given topic as well as links to their profile page or personal website for users in need of further information. Hence, it will be useful for stakeholders to easily identify available expertise at an institution, e.g., to initiate collaborations, research transfer, or advanced training.

One of the current limitations of our approach is that not all researchers have a profile on a scholarly database like Google Scholar, which might make it difficult to retrieve their publications and derive the research areas in which they are active. Another limitation is that getting a PDF version of a publication is often not possible due to the paywalls used by many publishers. Hence, in conclusion, although our system already delivers promising results in its early stages of development, there is plenty of room and need for future work.

\bibliographystyle{ACM-Reference-Format}
\bibliography{transform}


\begin{thebibliography}{18}


\ifx \showCODEN    \undefined \def \showCODEN     #1{\unskip}     \fi
\ifx \showDOI      \undefined \def \showDOI       #1{#1}\fi
\ifx \showISBNx    \undefined \def \showISBNx     #1{\unskip}     \fi
\ifx \showISBNxiii \undefined \def \showISBNxiii  #1{\unskip}     \fi
\ifx \showISSN     \undefined \def \showISSN      #1{\unskip}     \fi
\ifx \showLCCN     \undefined \def \showLCCN      #1{\unskip}     \fi
\ifx \shownote     \undefined \def \shownote      #1{#1}          \fi
\ifx \showarticletitle \undefined \def \showarticletitle #1{#1}   \fi
\ifx \showURL      \undefined \def \showURL       {\relax}        \fi
\providecommand\bibfield[2]{#2}
\providecommand\bibinfo[2]{#2}
\providecommand\natexlab[1]{#1}
\providecommand\showeprint[2][]{arXiv:#2}

\bibitem[Askari et~al\mbox{.}(2023)]%
        {askari2023test}
\bibfield{author}{\bibinfo{person}{Arian Askari}, \bibinfo{person}{Mohammad Aliannejadi}, \bibinfo{person}{Evangelos Kanoulas}, {and} \bibinfo{person}{Suzan Verberne}.} \bibinfo{year}{2023}\natexlab{}.
\newblock \showarticletitle{A test collection of synthetic documents for training rankers: Chatgpt vs. human experts}. In \bibinfo{booktitle}{\emph{Proceedings of the 32nd ACM International Conference on Information and Knowledge Management}}. \bibinfo{pages}{5311--5315}.
\newblock


\bibitem[Dong et~al\mbox{.}(2017)]%
        {dong2017century}
\bibfield{author}{\bibinfo{person}{Yuxiao Dong}, \bibinfo{person}{Hao Ma}, \bibinfo{person}{Zhihong Shen}, {and} \bibinfo{person}{Kuansan Wang}.} \bibinfo{year}{2017}\natexlab{}.
\newblock \showarticletitle{A century of science: Globalization of scientific collaborations, citations, and innovations}. In \bibinfo{booktitle}{\emph{Proceedings of the 23rd ACM SIGKDD international conference on knowledge discovery and data mining}}. \bibinfo{pages}{1437--1446}.
\newblock


\bibitem[Dutta et~al\mbox{.}(2015)]%
        {dutta2015mod}
\bibfield{author}{\bibinfo{person}{Biswanath Dutta}, \bibinfo{person}{Durgesh Nandini}, {and} \bibinfo{person}{Gautam~Kishore Shahi}.} \bibinfo{year}{2015}\natexlab{}.
\newblock \showarticletitle{MOD: metadata for ontology description and publication}. In \bibinfo{booktitle}{\emph{International Conference on Dublin Core and Metadata Applications}}. \bibinfo{pages}{1--9}.
\newblock


\bibitem[Farooq(2023)]%
        {farooq2023knowledge}
\bibfield{author}{\bibinfo{person}{Rayees Farooq}.} \bibinfo{year}{2023}\natexlab{}.
\newblock \showarticletitle{Knowledge management and performance: a bibliometric analysis based on Scopus and WOS data (1988--2021)}.
\newblock \bibinfo{journal}{\emph{Journal of Knowledge Management}} \bibinfo{volume}{27}, \bibinfo{number}{7} (\bibinfo{year}{2023}), \bibinfo{pages}{1948--1991}.
\newblock


\bibitem[Guest et~al\mbox{.}(2013)]%
        {guest2013collecting}
\bibfield{author}{\bibinfo{person}{Greg Guest}, \bibinfo{person}{Emily~E Namey}, {and} \bibinfo{person}{Marilyn~L Mitchell}.} \bibinfo{year}{2013}\natexlab{}.
\newblock \bibinfo{booktitle}{\emph{Collecting qualitative data: A field manual for applied research}}.
\newblock \bibinfo{publisher}{Sage}.
\newblock


\bibitem[Lin(2024)]%
        {lin2024revolutionizing}
\bibfield{author}{\bibinfo{person}{Demiao Lin}.} \bibinfo{year}{2024}\natexlab{}.
\newblock \showarticletitle{Revolutionizing Retrieval-Augmented Generation with Enhanced PDF Structure Recognition}.
\newblock \bibinfo{journal}{\emph{arXiv preprint arXiv:2401.12599}} (\bibinfo{year}{2024}).
\newblock


\bibitem[Nandini and Shahi(2019)]%
        {nandini2019ontology}
\bibfield{author}{\bibinfo{person}{Durgesh Nandini} {and} \bibinfo{person}{Gautam~Kishore Shahi}.} \bibinfo{year}{2019}\natexlab{}.
\newblock \showarticletitle{An ontology for transportation system}.
\newblock  (\bibinfo{year}{2019}).
\newblock


\bibitem[Nedumov and Kuznetsov(2019)]%
        {nedumov2019exploratory}
\bibfield{author}{\bibinfo{person}{Ya~R Nedumov} {and} \bibinfo{person}{Sergei~D Kuznetsov}.} \bibinfo{year}{2019}\natexlab{}.
\newblock \showarticletitle{Exploratory search for scientific articles}.
\newblock \bibinfo{journal}{\emph{Programming and Computer Software}}  \bibinfo{volume}{45} (\bibinfo{year}{2019}), \bibinfo{pages}{405--416}.
\newblock


\bibitem[Okey et~al\mbox{.}(2023)]%
        {okey2023investigating}
\bibfield{author}{\bibinfo{person}{Ogobuchi~Daniel Okey}, \bibinfo{person}{Ekikere~Umoren Udo}, \bibinfo{person}{Renata~Lopes Rosa}, \bibinfo{person}{Demostenes~Zegarra Rodr{\'\i}guez}, {and} \bibinfo{person}{Jo{\~a}o~Henrique Kleinschmidt}.} \bibinfo{year}{2023}\natexlab{}.
\newblock \showarticletitle{Investigating ChatGPT and cybersecurity: A perspective on topic modeling and sentiment analysis}.
\newblock \bibinfo{journal}{\emph{Computers \& Security}}  \bibinfo{volume}{135} (\bibinfo{year}{2023}), \bibinfo{pages}{103476}.
\newblock


\bibitem[Ortiz~Vivar et~al\mbox{.}(2022)]%
        {ortiz2022redi}
\bibfield{author}{\bibinfo{person}{Jos{\'e} Ortiz~Vivar}, \bibinfo{person}{Jos{\'e} Segarra}, \bibinfo{person}{Boris Villaz{\'o}n-Terrazas}, {and} \bibinfo{person}{V{\'\i}ctor Saquicela}.} \bibinfo{year}{2022}\natexlab{}.
\newblock \showarticletitle{REDI: Towards knowledge graph-powered scholarly information management and research networking}.
\newblock \bibinfo{journal}{\emph{Journal of Information Science}} \bibinfo{volume}{48}, \bibinfo{number}{2} (\bibinfo{year}{2022}), \bibinfo{pages}{167--181}.
\newblock


\bibitem[Perrier et~al\mbox{.}(2017)]%
        {perrier2017research}
\bibfield{author}{\bibinfo{person}{Laure Perrier}, \bibinfo{person}{Erik Blondal}, \bibinfo{person}{A~Patricia Ayala}, \bibinfo{person}{Dylanne Dearborn}, \bibinfo{person}{Tim Kenny}, \bibinfo{person}{David Lightfoot}, \bibinfo{person}{Roger Reka}, \bibinfo{person}{Mindy Thuna}, \bibinfo{person}{Leanne Trimble}, {and} \bibinfo{person}{Heather MacDonald}.} \bibinfo{year}{2017}\natexlab{}.
\newblock \showarticletitle{Research data management in academic institutions: A scoping review}.
\newblock \bibinfo{journal}{\emph{PLoS One}} \bibinfo{volume}{12}, \bibinfo{number}{5} (\bibinfo{year}{2017}), \bibinfo{pages}{e0178261}.
\newblock


\bibitem[Richardson(2007)]%
        {richardson2007beautiful}
\bibfield{author}{\bibinfo{person}{Leonard Richardson}.} \bibinfo{year}{2007}\natexlab{}.
\newblock \bibinfo{title}{Beautiful soup documentation}.
\newblock
\newblock


\bibitem[Ries(2011)]%
        {ries2011lean}
\bibfield{author}{\bibinfo{person}{Eric Ries}.} \bibinfo{year}{2011}\natexlab{}.
\newblock \bibinfo{booktitle}{\emph{The lean startup: How today's entrepreneurs use continuous innovation to create radically successful businesses}}.
\newblock \bibinfo{publisher}{Currency}.
\newblock


\bibitem[Salaba and Chan(2023)]%
        {salaba2023cataloging}
\bibfield{author}{\bibinfo{person}{Athena Salaba} {and} \bibinfo{person}{Lois~Mai Chan}.} \bibinfo{year}{2023}\natexlab{}.
\newblock \bibinfo{booktitle}{\emph{Cataloging and classification: an introduction}}.
\newblock \bibinfo{publisher}{Rowman \& Littlefield}.
\newblock


\bibitem[Schopf et~al\mbox{.}(2023)]%
        {schopf2023knowledge}
\bibfield{author}{\bibinfo{person}{Tim Schopf}, \bibinfo{person}{Nektrios Machner}, {and} \bibinfo{person}{Florian Matthes}.} \bibinfo{year}{2023}\natexlab{}.
\newblock \showarticletitle{A Knowledge Graph Approach for Exploratory Search in Research Institutions}.
\newblock \bibinfo{journal}{\emph{arXiv preprint arXiv:2311.15688}} (\bibinfo{year}{2023}).
\newblock


\bibitem[Schuetzenmeister(2010)]%
        {schuetzenmeister2010university}
\bibfield{author}{\bibinfo{person}{Falk Schuetzenmeister}.} \bibinfo{year}{2010}\natexlab{}.
\newblock \showarticletitle{University research management: An exploratory literature review}.
\newblock  (\bibinfo{year}{2010}).
\newblock


\bibitem[Shahi and Kana~Tsoplefack(2022)]%
        {shahi2022mitigating}
\bibfield{author}{\bibinfo{person}{Gautam~Kishore Shahi} {and} \bibinfo{person}{William Kana~Tsoplefack}.} \bibinfo{year}{2022}\natexlab{}.
\newblock \showarticletitle{Mitigating Harmful Content on Social Media Using an Interactive User Interface}. In \bibinfo{booktitle}{\emph{International Conference on Social Informatics}}. Springer, \bibinfo{pages}{490--505}.
\newblock


\bibitem[Strobel et~al\mbox{.}(2024)]%
        {strobel2024exploring}
\bibfield{author}{\bibinfo{person}{Gero Strobel}, \bibinfo{person}{Leonardo Banh}, \bibinfo{person}{Frederik M{\"o}ller}, {and} \bibinfo{person}{Thorsten Schoormann}.} \bibinfo{year}{2024}\natexlab{}.
\newblock \showarticletitle{Exploring generative artificial intelligence: A taxonomy and types}.
\newblock  (\bibinfo{year}{2024}).
\newblock


\end{thebibliography}

\balance
\end{document}